\documentclass{article}
\begin{document}
\large

\begin{center} {\Large \bf Path Integral Quantization of Dual Abelian Gauge Theory} \\
\vspace{2ex}
Mark S. Swanson\footnote{E-Mail: Swanson@UConnVM.UConn.Edu} \\

{\it Department of Physics, University of Connecticut, Stamford, Connecticut 06901}
\end{center}

\begin{abstract}
The path integral for 3+1 abelian gauge theory is rewritten in terms of a real antisymmetric field allowing a dual action that couples the electric and magnetic currents to the photon and each other in a gauge invariant manner.  Standard perturbative abelian quantum electrodynamics reemerges when the monopole current vanishes. For certain simple relationships between the monopole current and the electric current, the altered photon propagator can exhibit abelian charge confinement or develop mass, modeling effects believed to be present in non-abelian theories. 
\end{abstract}

\vspace{3ex}

Since the classic work of Dirac \cite{Dirac} magnetic monopoles have been studied in a variety of contexts.  Retaining the usual gauge field structure requires singular gauge field configurations in the presence of magnetic charge.  Dirac was the first to show that the singularities --- Dirac ``strings'' --- are not physical as long as the electric and magnetic charges satisfy the quantization condition $eg = \frac{1}{2} n$ and its generalization, where $n$ is an integer.  Wu and Yang \cite{WuYang} showed the relationship of the Dirac monopole to a U(1) principal bundle with the base manifold given by a sphere around the monopole, with single-valuedness of the transition function responsible for the quantization condition.  Beginning with the work of 't~Hooft and Polyakov \cite{tHooftPolyakov} much attention has been focused on BPS monopoles in non-abelian gauge theories \cite{Harvey}.  An important result regarding the effects of magnetic monopoles was the demonstration by Seiberg and Witten \cite{SeibergWitten} that ${\cal N} =2$ supersymmetric QCD, a field theory exhibiting duality, breaks chiral symmetry and confines through the presence of monopole-antimonopole pairs in the ground state. It has been widely considered that monopoles may play a similar role in standard QCD \cite{tHooftBruckmann}.

The current value of the Parker bound \cite{Freese} indicates that free magnetic monopoles are rare, if indeed they exist at all. However, their inclusion in quantum processes has many intriguing aspects.  Zwanziger \cite{Zwanziger} addressed this problem by modifying the standard abelian action to include a dual gauge field. Gamsberg and Milton \cite{GamsbergMilton} analyzed the current--current interaction term, first proposed by Schwinger \cite{Schwinger}, using a path integral over the gauge field. In non-abelian theories monopole effects are evaluated by perturbing around the monopole field configuration, in effect treating the monopole as a classical solution \cite{Coleman}.  Making this approach quantum mechanically consistent for an arbitrary set of monopoles requires integrating the associated path integral over the monopole moduli space, and much effort has gone into solving this difficult problem \cite{PaisSchroers}.  

The purpose of this paper is to derive the path integral for self-dual 3+1 abelian gauge theory,  allowing both a magnetic current $J$ and an electric current $\jmath$.  For the purposes of this paper, self-duality means that the path integral is invariant under the transformation $J \rightarrow - \jmath$ and $\jmath \rightarrow J$.  When coupled with the transformations of the electric and magnetic fields ${\bf E} \rightarrow {\bf B}$ and ${\bf B} \rightarrow - {\bf E}$, this is an invariance of the classical Maxwell's equations in the presence of magnetic charge \cite{Jackson}. For the case that the magnetic current $J$ is globally proportional to the usual electric current $\jmath$, there exists a duality transformation that eliminates the magnetic current, returning Maxwell's equations to their standard form.  However, the standard Maxwell action for the gauge field, $\frac{1}{2} ( {\bf E}^2 - {\bf B}^2 )$, is not invariant under a duality transformation of the electric and magnetic fields. As a result, a path integral quantization employing the standard Maxwell action cannot manifest duality and does not represent quantized abelian gauge theory in the presence of magnetic charge in a manner consistent with Maxwell's equations.

The path integral presented in this paper is an extension of the standard abelian gauge theory  to one that possesses exact symmetry under the operation ${\bf E} \rightarrow {\bf B}$ and ${\bf B} \rightarrow - {\bf E}$.  Inclusion of this symmetry gives rise to a coupling between $\jmath$ and $J$ that manifests duality and that is consistent with the results of Zwanziger \cite{Zwanziger} and Gamsberg and Milton \cite{GamsbergMilton}. In addition to allowing distinct magnetic and electric currents, the resulting path integral also allows simple models of a  monopole condensate around the electric charges consistent with the antiscreening nature of non-abelian magnetic charges \cite{Goebel}. For simple cases it will be seen that confinement of electric charge or mass generation can occur due to monopole effects. The model therefore gives a simple abelian laboratory for examining the effects of magnetic coupling thought present in non-abelian theories.

In the standard formulation of abelian gauge theory the Maxwell two-form {\bf F} is derived from the gauge field one-form {\bf A} by exterior differentiation ${\bf F} = {\rm d}{\bf A}$.  This automatically implies  that ${\rm d}{\bf F} = {\rm d}^2 {\bf A} = 0$ for non-singular configurations.  However, the components of ${\rm d}{\bf F}$ are given by $\epsilon_{\mu \nu \rho \sigma} \partial^\nu F^{\rho \sigma}$, and these are the components of the magnetic monopole current.  Retaining the gauge structure of the Maxwell two-form --- essential to defining minimal coupling to the matter fields --- therefore requires singular configurations such that ${\textrm d}^2 {\bf A} \not = 0$ for at least some set of points.  Direct integration over such configurations in a path integral is problematic since the measure is almost always defined by the modes determined from the elliptic operators present in the Wick-rotated Lagrangian.  In the case of an unbroken abelian gauge field on non-compact Euclidean space this reduces to integrating over the Fourier modes, where the gauge field $A$ is written $A_\mu (x) = \int_k \tilde{A}_\mu (k) \, e^{ikx}$.  Coupled with the restrictions $\tilde{A}_\mu ( -k) = \tilde{A}_\mu (k)$ and $\tilde{A}^*_\mu (k) = \tilde{A}_\mu (k)$, this decomposition gives a real gauge field, $A^*_\mu (x) = A_\mu (x)$. This decomposition is {\it a priori} non-singular since $\left[ \partial_\mu , \partial_\nu \right] A_\rho ( x ) = i \int_k \left[ k_\mu , k_\nu \right] \tilde{A}_\rho (k) \, e^{ikx} = 0$. If such a measure is used, integration over monopole moduli space is the only viable method for including monopole contributions in the quantized theory, and this increases the difficulty in evaluating virtual or production processes involving the monopole matter fields.  

An alternative method for including magnetic monopoles in the quantized theory is to expand the space of functions being integrated consistent with gauge invariance.  This is similar to Zwanziger's approach \cite{Zwanziger} to monopole-charge coupling. The first step is to rewrite the standard abelian path integral with only an electric current $\jmath_\mu$ present in terms of a gauge-invariant integration over a real antisymmetric field $\varphi_{\mu \nu} (x) = i \int_k \, \tilde{\varphi}_{\mu \nu} (k) \, e^{ikx}$, whose Fourier transform satisfies $ \tilde{\varphi}_{\mu \nu} ( - k ) = - \tilde{\varphi}_{\mu \nu} ( k )$ and $\tilde{\varphi}^*_{\mu \nu} (k) = \tilde{\varphi}_{\mu \nu} (k) $. These restrictions insure that $\varphi$ is real as well as satisfying the necessary PT transformation property $\varphi_{\mu \nu} ( - x ) = - \varphi_{\mu \nu} ( x )$.  Only the gauge field part of the action and measure will be evaluated, so that the conserved electric current $\jmath_\mu$ can be either quantum mechanical or an external classical source. However, gauge invariance can be realized only if the matter action is present. The decomposition for $\jmath$ takes the form $\jmath_\mu (x) = \int_k \tilde{\jmath}_\mu (k) e^{ikx}$ with the restriction $k_\mu \tilde{\jmath}^{\, \mu} = 0$ to satisfy current conservation.  Denoting $F_{\mu \nu} = \partial_\mu A_\nu - \partial_\nu A_\mu$ and using an obvious notation it follows that 
\begin{eqnarray}
\label{Rewrite}
Z[\jmath] & = & \int \left[ {\rm d} \varphi \right] \left[ {\rm d} A \right] \exp i \int \left(  - \frac{1}{4} \varphi \cdot \varphi + \frac{1}{2} i \varphi \cdot F - \jmath \cdot A  \right)  
\nonumber
\\ & = &  \int \left[ {\rm d} A \right] \exp i \int \left( - \frac{1}{4} F \cdot F  - \jmath \cdot A \right) \; ,
\end{eqnarray} 
where $\left[ \cdots \right]$ stands for a product of normalization factors as well as the actual measure.  Result (\ref{Rewrite}) follows by performing a Wick rotation \cite{WickRotation} and the subsequent Gaussian integrations using the result $\int {\rm d}x \, \exp ( - {\small \frac{1}{2}} \alpha^2 x^2 \pm i \beta x ) = \exp (-  {\small \frac{1}{2}} \beta^2)$.   Inclusion of a theta term will be considered elsewhere.

The gauge field $A$ can be integrated from the first path integral in (\ref{Rewrite}) since the action is linear in the Fourier mode $\tilde{A}_\mu$.  The resulting momentum space path integral is given by
\begin{equation}
\label{NewQED}
Z[\jmath] = \int \left[ {\rm d} \tilde{\varphi} \,  \delta( k \cdot \tilde{\varphi} + \tilde{\jmath} \,) \right]  \, \exp \left( i \int_k - {\small \frac{1}{4}} \tilde{\varphi} \cdot \tilde{\varphi} \right) \; .
\end{equation}
The path integral of (\ref{NewQED}) has six degrees of freedom in the measure and four constraints representing the two matter-coupled Maxwell equations.  In effect (\ref{NewQED}) is a parent theory to the standard formulation of abelian electrodynamics. However, (\ref{NewQED}) places ${\bf B}$ and ${\bf E}$ on an equal footing in the measure, allowing a broader class of magnetic field configurations to be included in the integrations. 

The path integral (\ref{NewQED}) will now be extended to include non-trivial monopole currents $J_\mu$.  The first step will be to deduce the need for constraints reflecting the dual Maxwell equations. The motivation for this, apart from developing a dual path integral, is two-fold. The first comes from evaluating the  the case $J_\mu = 0$ by giving the variable $\tilde{\varphi}$ the decomposition
\begin{equation}
\label{NoMonopole}
\tilde{\varphi}_{\mu \nu} (k)  =   k_\mu ( \tilde{A}_\nu  + \xi^\lambda_\nu a^\lambda )  - k_\nu (  \tilde{A}_\mu  + \xi^\lambda_\mu a^\lambda ) \; ,
\end{equation}
where the $\xi^\lambda$ are the usual pair of transverse polarization vectors satisfying $k_\mu \xi^{\mu \lambda} = 0$ and $\xi_\mu^\lambda \xi^{\mu \lambda^\prime} = - \delta^{\lambda \lambda^\prime}$.  This decomposition implements the desired constraint $\tilde{J}_\mu = \frac{1}{2} \epsilon_{\mu \nu \rho \sigma} k^\nu \tilde{\varphi}^{\rho \sigma} = 0$. It has the usual gauge invariance under $\tilde{A}_\mu \rightarrow \tilde{A}_\mu + k_\mu \Lambda (k)$, as well as an additional invariance under the simultaneous operations $\tilde{A}_\mu \rightarrow \tilde{A}_\mu + \xi^\lambda_\mu f^\lambda$ and $a^\lambda \rightarrow a^\lambda - f^\lambda$.  The latter symmetry insures that the $a^\lambda$ are transverse ghosts and can be gauged from the theory, just as the usual gauge symmetry insures that the Lorentz gauge ghost $k_\mu \tilde{A}^\mu$ can be gauged from the theory.  These symmetries create three gauge volumes in (\ref{NewQED}) as opposed to only one. The decomposition and gauge-fixing recreates the standard Lorentz gauge action. It also reduces the measure to the standard Lorentz gauge integration over $\tilde{A}_\mu$ since trivial Faddeev--Popov determinants of unity result from imposing the gauge condition $\prod_\lambda \delta(a^\lambda)$.  The constraint takes the form $\delta( k^2 \tilde{A}_\mu + \tilde{\jmath}_\mu )$, and this reproduces the standard Lorentz gauge photon propagator when the $\tilde{A}_\mu$ variables are integrated. 

However, there is a critical subtlety revealed by the fact that the Jacobian of the transformation (\ref{NoMonopole}) vanishes.  An evaluation of the Jacobian can be accomplished by explicitly breaking the symmetries of (\ref{NoMonopole}) through a redefinition of the expansion:
\begin{equation}
\label{Regulate}
\tilde{\varphi}_{\mu \nu, \alpha} (k)  =   k_\mu (1 - \alpha )( \tilde{A}_\nu  + \xi^\lambda_\nu a^\lambda )  - \, k_\nu \, (  \tilde{A}_\mu  + \xi^\lambda_\mu a^\lambda ) +  \alpha \epsilon_{\mu \nu \rho \sigma} k^\rho  \tilde{A}^\sigma \; .
\end{equation}
The Jacobian of (\ref{Regulate}) is given by $\alpha^4 k^4 k_0 |\vec{k}|$, showing the singular nature of the limit $\alpha \rightarrow 0$.  Of course, these singularities are not present in the standard abelian path integral, showing that the correct measure for the extended path integral must contain other factors to cancel them.  These Jacobian singularities can be regulated if the four dual constraint equations $\delta ( {\small \frac{1}{2}} \epsilon_{\mu \nu \rho \sigma} k^\nu \tilde{\varphi}^{\rho \sigma} + \alpha \tilde{\jmath}_\mu )$ are present in the measure.  These four constraints correspond to the dual Maxwell equations ${\rm d}F = 0$ after an infinitesimal duality transformation.  Showing that this eliminates the singularities requires the limiting procedure of (\ref{Regulate}).  Expanding the dual constraints using (\ref{Regulate}) gives
\begin{eqnarray}
\label{Cancellation}
& &   \lim_{\alpha \rightarrow 0} \alpha \, \delta( {\small \frac{1}{2}} \epsilon_{\mu \nu \rho \sigma} k^\nu \tilde{\varphi}^{\rho \sigma}_\alpha + \alpha \tilde{\jmath}_\mu ) \nonumber \\
& & = \lim_{\alpha \rightarrow 0} \alpha \, \delta \left( \alpha ( k^2 \tilde{A}_\mu - k_\mu \, k \cdot \tilde{A} + \tilde{\jmath}_\mu ) \right) \nonumber \\
& & = \delta ( k^2 \tilde{A}_\mu - k_\mu \, k \cdot \tilde{A} + \tilde{\jmath}_\mu )
\end{eqnarray}
so that all four singularities from the Jacobian are canceled and the limit $\alpha \rightarrow 0$ is well-defined. In the presence of (\ref{Cancellation}) the other four constraints become $\lim_{\alpha \rightarrow 0} \prod_\mu \delta( k^\nu \tilde{\varphi}_{\nu \mu , \alpha} + \tilde{\jmath}_\mu \,) = \prod_\mu \delta ( k^2 \xi^\lambda_\mu a^\lambda ) = \delta( a^1 ) \, \delta ( a^2 ) \, \delta^2 ( 0 ) / k^4$. As a result, the two gauge volumes for the $a^\lambda$ have been factorized in the form $\delta^2(0)$ and the previously discussed constraints on $a^\lambda$ have emerged. Enforcing the Lorentz gauge constraint $k \cdot \tilde{A} = 0$ removes the final gauge volume and allows the $\tilde{A}_\mu$ variables to be integrated. Using the fact that the action is given by $- \frac{1}{4} \tilde{\varphi}^2 =  - \frac{1}{2} k^2 \tilde{A}^2$, the usual Minkowski momentum space form for the Lorentz gauge path integral is obtained:
\begin{equation}
\label{LorentzGauge}
Z [\jmath] = \exp \left( -  i \int_k  \frac{ \tilde{\jmath}^{\, 2} }{2 k^2} \right) \; .
\end{equation}

A second compelling reason to include the dual constraints is to realize gauge invariance in the monopole sector of the theory for the case $\tilde{J}$ is non-zero.  Two separate gauge invariances are required in the presence of a non-trivial magnetic charge, since for such a case the electric charge of the monopole field must be different from that of the currently observed fundamental electric charge, and the matter sector of the theory cannot be invariant under a single gauge transformation.  The second constraint solves this problem since it can be written 
\begin{equation}
\label{Exponentiation}
 \prod_k \delta( \frac{1}{2} \epsilon_{\mu \nu \rho \sigma} k^\nu \tilde{\varphi}^{\rho \sigma} - \tilde{J}_\mu \,)  = \int [{\rm d}\tilde{B}] 
\, \exp i \int_k  \left( \frac{1}{2} \epsilon_{\mu \nu \rho \sigma} \tilde{B}^\mu k^\nu \tilde{\varphi}^{\rho \sigma} - \tilde{B}^\mu \tilde{J}_\mu  \right) 
\end{equation}
and absorbed into the action of the path integral where $\tilde{B}_\mu$ provides the second gauge invariance required to couple the monopole matter action. In this regard, it is critical that the first term in the exponential of (\ref{Exponentiation}) and the measure $[{\rm d} \tilde{B}]$ are both invariant under the gauge transformation $\tilde{B}_\mu \rightarrow \tilde{B}_\mu + k_\mu \Lambda^\prime$.

While the constrained measure now implements the necessary gauge invariance and reflects duality, the Maxwell action does not respect the latter. Using the two constraints and the properties of the Levi-Civita symbol, it is straightforward to show that $ - \frac{1}{4} \tilde{\varphi}^2 = ( \tilde{J}^{\, 2} - \tilde{\jmath}^{\, 2} )/2k^2$.  Besides lacking duality symmetry, the standard Maxwell action generates no coupling between the two currents and an incorrect coupling between the magnetic current and the photon.  It is therefore correct only for the case $\tilde{J} = 0$. 

The question of the correct action in the presence of both electric and magnetic currents can be finessed by starting with the situation where no currents are present. Performing the translation $\varphi_{\mu \nu} \rightarrow \varphi_{\mu \nu} + m_{\left[ \mu \right.} \tilde{\jmath}_{\left. \nu \right]}$, where the momentum space ``string''  $m_\mu$ satisfies $k^\mu m_\mu = 1$, adds the electric current $\tilde{\jmath}_\mu$ to the first constraint and the magnetic current $\tilde{J}_\mu =  - \epsilon_{\mu \nu \rho \sigma} k^\nu m^\rho \tilde{\jmath}^\sigma$ to the second constraint.  This translation also induces the additional terms $ - \tilde{\varphi}^{\mu \nu} m_\mu \tilde{\jmath}_\nu - \frac{1}{2} \left[ \tilde{\jmath}^{\, 2} m^2 - (\tilde{\jmath} \cdot m )^2 \right]$ in the action.  For this form of the magnetic current, it can be shown that  $\frac{1}{2} \left[ \tilde{\jmath}^{\, 2} m^2 - (\tilde{\jmath} \cdot m)^2  \right] = ( \tilde{\jmath}^{\, 2} - \tilde{J}^{\, 2})/2k^2 =  \frac{1}{4} \tilde{\varphi}^2$, so that the total action after translation can be written $S = - \frac{1}{2} \tilde{\varphi}^2 +  m_\mu \tilde{\varphi}^{\mu \nu} k^\rho \tilde{\varphi}_{\rho \nu}$. Subtracting $S$ from the original path integral action defines a self-dual path integral for the case that both currents are present:
\begin{equation}
\label{FinalForm}
Z[ \jmath, J ]  = \int \left[ {\rm d} \tilde{\varphi} \,  \delta( k^\nu \tilde{\varphi}_{\nu \mu} + \tilde{\jmath}_\mu \,) \, \delta ( \frac{1}{2} \epsilon_{\mu \nu \rho \sigma} k^\nu \tilde{\varphi}^{\rho \sigma} - \tilde{J}_\mu ) \, \right] \exp i \int_k  \left( \frac{1}{4} \tilde{\varphi}_{\mu \nu} \tilde{\varphi}^{\mu \nu} - k^\mu \tilde{\varphi}_{\mu \nu} m_\rho \tilde{\varphi}^{ \rho \nu}   \right) \; .
\end{equation}
The action in (\ref{FinalForm}) is now invariant under the transformation $\tilde{\varphi}_{\mu \nu} \rightarrow  \frac{1}{2} \epsilon_{\mu \nu \rho \sigma} \tilde{\varphi}^{\rho \sigma}$,  realizing the duality symmetry ${\bf E} \rightarrow {\bf B}$ and ${\bf B} \rightarrow - {\bf E}$. It is straightforward to show that the constraints in the measure cause the path integral action to localize onto $ \frac{1}{4} \tilde{\varphi}_{\mu \nu} \tilde{\varphi}^{\mu \nu} - k^\mu \tilde{\varphi}_{\mu \nu} m_\rho \tilde{\varphi}^{ \rho \nu} = - \frac{1}{2} (\tilde{\jmath}^{\, 2} + \tilde{J}^{\, 2} - 2 \epsilon_{\mu \nu \rho \sigma} \tilde{\jmath}^{\, \mu} k^\nu m^\rho \tilde{J}^{\, \sigma})/ k^2$, resulting in the correct coupling to the photon propagator for both electric and magnetic currents as well as the interaction term first proposed by Schwinger \cite{Schwinger}.  {\it This result will be true independent of the mode decompositions for $\tilde{\varphi}$ up to a gauge-fixing term}, and is manifestly invariant under the duality transformations of the currents.

Both constraints and the modified action are required in (\ref{FinalForm}) to enforce Maxwell's equations in the presence of both $\tilde{\jmath}$ and $\tilde{J}$, as well to prevent singularities in the limit that the magnetic current vanishes.  It is important to note that there are only six real degrees of freedom if both currents are present since both currents are conserved. The degrees of freedom in the currents therefore match the six degrees of freedom in the $\tilde{\varphi}$ measure, showing that the additional degrees of freedom present in the extended path integral are essential to incorporating magnetic charge.  The choice of mode expansion for $\tilde{\varphi}$ is determined for consistency with the type of currents present and the requirements of gauge invariance. For example, the mode decomposition (\ref{NoMonopole}) is consistent with $\tilde{J}_\mu = 0$ and reduces the action to its usual quadratic form.  Two additional cases will now be considered.

{\it Case i: Distinct $J$}.  For the case that there are two distinct currents, the most convenient mode expansion is given by emulating Zwanziger and using a dual expansion:
\begin{equation}
\label{DualExpansion}
\tilde{\varphi}_{\mu \nu} = k_{\left[ \mu \right.} \tilde{A}_{\left. \nu \right]} +  \epsilon_{\mu \nu \rho \sigma} k^\rho \tilde{B}^\sigma \; .
\end{equation}
In order to reduce the eight variables of (\ref{DualExpansion}) to the six associated with $\tilde{\varphi}$ it is necessary to introduce two constraints on the $\tilde{A}$ and $\tilde{B}$ variables.  These constraints are possible since (\ref{DualExpansion}) is invariant under two separate gauge symmetries: $\tilde{A}_\mu \rightarrow \tilde{A}_\mu + k_\mu \Lambda$ and  $\tilde{B}_\mu \rightarrow \tilde{B}_\mu + k_\mu \Lambda^\prime$.  The two gauge invariances can be used to enforce the Lorentz conditions $k \cdot \tilde{A} = k \cdot \tilde{B} = 0$.  For this choice, the Jacobian associated with (\ref{DualExpansion}) is non-singular, given by $k^4 |\vec{k}|^2$.  The action reduces to
\begin{equation}
\label{DualAction}
\frac{1}{4} \tilde{\varphi}^2 - k^\mu \tilde{\varphi}_{\mu \nu} m_\rho \tilde{\varphi}^{ \rho \nu}  = - \frac{1}{2} k^2 ( \tilde{A}^2 + \tilde{B}^2 ) -  k^2 \epsilon_{\mu \nu \rho \sigma} \tilde{A}^\mu k^\nu m^\rho \tilde{B}^\sigma \; .
\end{equation}
The coupling between $\tilde{A}$ and $\tilde{B}$ respects the dual gauge invariance necessary for non-trivial magnetic currents.  The constraints reduce to $\delta( k^2 \tilde{A}_\mu + \tilde{\jmath}_\mu)$ and $\delta( k^2 \tilde{B}_\mu - \tilde{J}_\mu )$, and these are consistent with the two Lorentz conditions, since $k_\mu \tilde{A}^\mu \rightarrow - k_\mu \tilde{\jmath}^\mu / k^2  =  0 $. Integrating the $\tilde{A}$ and $\tilde{B}$ variables gives the normalized functional consistent with the previously discussed localization result:
\begin{equation}
\label{FinalFunctional}
Z[ \jmath, J] = \exp - i \int_k \left( \frac{\tilde{\jmath}^{\, 2}}{2k^2} + \frac{\tilde{J}^{\, 2}}{2k^2} -  \epsilon_{\mu \nu \rho \sigma} \frac{\tilde{\jmath}^{\, \mu}}{k^2} k^\nu m^\rho \tilde{J}^{\, \sigma} \right) \; .
\end{equation}
Gamsberg and Milton \cite{GamsbergMilton} have shown that the interaction term $\epsilon_{\mu \nu \rho \sigma} \tilde{\jmath}^{\, \mu} k^\nu m^\rho \tilde{J}^{\, \sigma}/ k^2$ gives rise to non-perturbative scattering cross-sections between electric and magnetic charges that are independent of the string variable $m_\mu$ as long as the Dirac quantization condition is satisfied.

{\it Case ii: $\jmath$ as Monopole Source}.  Within the framework of (\ref{FinalForm}) it is possible to model a system where the current $\jmath$ also serves as a source for monopoles, so that the fundamental charges also have a monopole charge. The following form for the monopole current, $\tilde{J}_\mu  = f(k) \, \tilde{\jmath}_\mu$, where $f$ is a dimensionless function of $k$, can be evaluated by the expansion $\tilde{\varphi}_{\mu \nu} (k)  =   k_\mu ( \tilde{A}_\nu  + \xi^\lambda_\nu a^\lambda )  - k_\nu (  \tilde{A}_\mu  + \xi^\lambda_\mu a^\lambda ) + f(k) \epsilon_{\mu \nu \rho \sigma} k^\rho A^\sigma$.   The abelian theory then gives a simple context to evaluate the effects of such a coupling on the gauge field propagator.  Choosing $f(k) = \lambda/k$, where $\lambda$ has units of inverse length, gives rise to a monopole density $e \lambda / 4 \pi r^2$ around a stationary point charge $e$ at the origin, in effect modeling a monopole condensate around the charge.  It follows from (\ref{FinalFunctional}) that the Lorentz gauge photon propagator is altered to 
\begin{equation}
\label{ConfiningPropagator}
\Delta_{\mu \nu} (x - y)  = - i \int_k g_{\mu \nu} \left( \frac{1}{k^2} + \frac{\lambda^2}{k^4}  \right) e^{ik ( x - y ) } \;.
\end{equation}
A single charge would be associated with a divergent energy, since the monopole number associated with $\lambda/4 \pi r^2$ diverges. However, the Euclidean form for (\ref{FinalFunctional}), coupled with (\ref{ConfiningPropagator}), shows that two oppositely charged classical point particles separated by the distance $R$ have a well-defined energy  
\begin{equation}
\label{ConfiningPotential}
E = - \frac{e^2}{4 \pi R} + \frac{e^2 \lambda^2 R}{4 \pi} \; ,
\end{equation}
demonstrating a simple confining potential for electric charge in addition to the usual Coulomb potential. 

Choosing the monopole current function to have the form $f(k) = M/\sqrt{k^2 - M^2}$, where $M$ is a mass scale, causes the photon propagator to be modified to
\begin{equation}
\label{MassivePropagator}
\Delta_{\mu \nu} (x - y)  = - i \int_k \frac{g_{\mu \nu}}{k^2 - M^2} e^{ik ( x - y ) } \; ,
\end{equation}
so that the monopole coupling gives rise to a massive propagator. Both effects are similar to those believed to be manifested in QCD \cite{tHooftBruckmann}.


\begin{thebibliography}{99}
\bibitem{WickRotation} Because the usual prescription $k_0 \rightarrow i k_0$ causes ${\rm d}^4k \rightarrow i {\rm d}^4k$, the Wick rotation of the Fourier transforms requires $\tilde{A}_0 \rightarrow \tilde{A}_0$, $\tilde{A}_j \rightarrow - i \tilde{A}_j$, $\tilde{\varphi}_{0j} \rightarrow \tilde{\varphi}_{0j}$, $\tilde{\varphi}_{jk} \rightarrow -i \tilde{\varphi}_{jk}$, $\tilde{\jmath}_0 \rightarrow -i \tilde{\jmath}_0$,  and $\tilde{\jmath}_k \rightarrow \tilde{\jmath}_k$.
\end{thebibliography}
\end{document}